\documentclass[12pt]{article}
\usepackage{graphicx}
\usepackage[cp1251]{inputenc}
\usepackage{epsfig}
\usepackage[english]{babel}

\date{}
\def\b{\begin{equation}}
\def\e{\end{equation}}
\def\bee{\begin{enumerate}}
\def\eee{\end{enumerate}}
\def\be{\begin{vmatrix}}
\def\ee{\end{vmatrix}}

\begin{document}
\setcounter{page}{1}
\bibliographystyle{unsrt2}
\pagestyle{plain}

\title{\bf Longitudinal Magnetization and specific heat of the anisotropic Heisenberg antiferromagnet
on
 Honeycomb lattice }
\author{F. Azizi, H. Rezania\thanks{Corresponding author. Tel./fax: +98 831 427 4556.
E-mail: rezania.hamed@gmail.com}}
\maketitle{\centerline{Department of Physics, Razi University,
Kermanshah, Iran}

\begin{abstract}
We study the effects of longitudinal magnetic field and temperature
 on the thermodynamic properties
 of two dimensional Heisenberg antiferromagnet on the honeycomb lattice in the presence of anisotropic Dzyaloshinskii-Moriya
interaction and next nearest neighbor coupling exchange constant.
In particular, the temperature dependence of specific heat
 have been investigated for various physical parameters in the model Hamiltonian.
Using a hard core bosonic representation, the behavior of thermodynamic properties has been studied by
means of excitation spectrum of mapped bosonic gas.
The effect of Dzyaloshinskii-Moriya interaction term on thermodynamic properties
has also been studied via the bosonic model by Green's function approach.
Furthermore we have studied the magnetic field dependence of specific heat and
magnetization for various anisotropy
parameters. At low temperatures,
the specific heat is found to be monotonically increasing with temperature for magnetic fields in the gapped field induced phase region.
We have found the magnetic field dependence of specific heat shows
 a monotonic decreasing behavior for various magnetic fields due
to increase of energy gap in the excitation spectrum.
Also we have studied the dependence of
magnetization on Dzyaloshinskii-Moriya interaction strength for different next nearest neighbor coupling constant.
 \end{abstract}
\vspace{0.5cm} {\it \emph{Keywords}}: Longitudinal magnetization; Heisenberg model; Green's
function.

\maketitle
\section{Introduction}
Quantum magnetism on geometrically two-dimensional frustrated spin
systems with $S=1/2$ have lately received massive attentions, due to
their potential for realizing the quantum spin liquid, a
magnetically disordered state which respects all the symmetries of
the systems, even at absolute zero temperature\cite{balent}. The
spin model, recently attracted many interests, is the Heisenberg
model with first and second antiferromagnetic exchange interaction
in honeycomb lattice. In sufficiently low spin systems, the quantum
mechanical zero point motion can forbid long range magnetic order
and produce a quantum spin liquid state, a correlated state that
breaks no symmetry and possesses topological properties, possibly
sustaining fractionalized excitations\cite{anderson1,
fazekas,liang,sachdev1,sandvik}. Although the triangular lattice was
first theoretically proposed by Anderson\cite{anderson1} as an ideal
benchmark to search for the quantum spin liquid, it was soon found
that the $S=1/2$ antiferromagnetic Heisenberg model on a triangular
lattice is magnetically ordered with a $120^{0}$ arrangement of the
spins. Despite the intense activity, only a small number of
triangular materials have been identified as possible candidates for
quantum spin liquid behavior such as th layered organic materials.
Hence, there is a need to find evidence for quantum spin liquid
behavior in more compounds. Honeycomb lattice materials have
attracted lots of attention in recent years due to their interesting
and poorly understood magnetic properties. Inorganic materials such
as Na$_{2}$ Co$_{2}$TeO$_{6}$\cite{lefan},
 BaM$_{2}$(XO$_{2}$)$_{2}$ (with X=As)\cite{martin},
Bi$_{3}$Mn$_{4}$O$_{12}$(NO$_{3}$)\cite{smirnova} and
In$_{3}$Cu$_{2}$VO$_{9}$\cite{yan} are examples of honeycomb lattice
antiferromagnets in which the magnitude of the spin varies from
$S=1/2$ in BaM$_{2}$(XO$_{4}$)$_{2}$ for M=Co to S=1 for M=Ni (with
X=As) and to S=3/2 in Bi$_{3}$Mn$_{4}$O$_{12}$(NO$_{3}$). It is
important then to understand theoretically the magnetic properties
of interacting localized moments on the frustrated honeycomb lattice
as has been previously done on triangular lattices. Although the
numerical evidence for a quantum spin liquid in the half-filled
Hubbard model on the honeycomb lattice\cite{meng} has been
questioned\cite{sorella}, exact diagonalization studies on the
$J_{1}-J_{2}$ Heisenberg model with $S=1/2$ have found evidence for
short range spin gapped phases for $J_{2}=0.3-0.35$ suggesting the
presence of a Resonance Valence Bond (RVB) state\cite{fouet}. In the
presence of external magnetic fields, finite temperature high
resolution spectroscopies such as inelastic neutron
scattering\cite{stone} and magnetic transport \cite{heidr} have
theoretically
 been calculated by dynamical
correlation functions of the Heisenberg model on honeycomb lattice. Specially, field
induced effects on the dynamical spin correlation function in low
dimensional quantum spin models have been attracting much interest
from theoretical and experimental point of view in recent
years
\cite{affleck,uimin,dimitriev}.
Spin-orbit coupling induces both symmetric and
antisymmetric anisotropic properties or Dzyaloshinskii-Moriya spin
anisotropy in the exchange coupling between nearest neighbour spins.
A Dzyaloshiskii-Moriya (DM) interaction with a DM vector
perpendicular to the layer produces an easy-plane spin anisotropic
Hamiltonian for honeycomb lattice. DM interaction breaks $SU(2)$
symmetry of the model and reduces it to $U(1)$ symmetry around the
$D$ vector. This DM interaction is believed to orient the spins in the 2D layer.
For applied large magnetic fields ($B$) along perpendicular to the plane, the Zeeman term overcomes the antiferromagnetic spin coupling
and the ground state
is a field induced ferromagnetic state with
gapped magnon excitations. Decreasing the magnetic field at the zero temperature,
the magnon gap vanishes at the critical field ($B_{c}$) and a spiral transverse magnetic
ordering develops.
At finite temperature ($T$)  and in the absence of
exchange anisotropy, the frustration leads to an incommensurate
spiral structure for transverse spin component below the critical line in the $B-T$ plane.

Hard core bosonic representation has been introduced to transform a spin
Hamiltonian to bosonic one whereby the excitation spectrum is obtained\cite{auerbach,mahan}.
This mapping between the bosonic gas and original spin model is valid provided to add the hard core
repulsion between particles in the bosonic Hamiltonian\cite{mahan}. An anisotropic exchange
interaction due to DM interaction adds a hopping term for bosonic particles to the main part of bosonic model Hamiltonian.
Such term leads to different universal behavior in addition to change of critical points
and thermodynamic properties. In our previous works, we have studied thermodynamic properties of low dimensional Heisenberg model Hamiltonian
such as Heisenberg chain and anisotropic spin ladders\cite{hamed11,hame22}

The goal of this work is to sort out the effect of magnetic field on the specific heat and longitudinal magnetization of two dimensional
 Heisenberg spin model on the honeycomb lattice in the field induced gapped spin-polarized phase.
Specially, we study the effects of Dzyaloshinskii-Moriya interaction strength and next nearest neighbor exchange constant on the
mentioned thermodynamic quantities as a function of temperature and
magnetic fields above critical field $B_{c}$. The spin model is
mapped to a bosonic one with infinitely strong repulsive short range
interaction \cite{mahan}. This infinite repulsion term preserves
SU(2) algebra of the spin model. Using Green's function approach,
the excitation spectrum of hard core Bosonic gas
 has been found within
Brueckner's approach \cite{gorkov}. In order to calculate
thermodynamic properties of the spin model Hamiltonian including
antisymmetric anisotropic Dzyaloshinskii-Moriya term, we have used one particle excitation spectrum
of hard core bosonic model Hamiltonian. In the last section we will
discuss and analyze our results to show how magnetic field and
DM interaction strength affect the temperature dependence of specific
heat and longitudinal magnetization. Furthermore the behavior of
these thermodynamic properties of this two dimensional Heisenberg model as a function
magnetic field for different values of DM interaction has
been studied. Also we have studied the effects  of next nearest neighbor coupling exchange constant on the behaviors of both longitudinal
magnetization and specific heat of this two dimensional spin model Hamiltonian.
\section{Theoretical formalism}
The antiferromagnetic Heisenberg model Hamiltonian on honeycomb lattice
 with anisotropic Dzyaloshinskii-Moriya interaction in the presence of Zeeman term
is defined by
\begin{eqnarray}
H=J\sum_{\langle ij\rangle}{\bf S}_{i}.{\bf S}_{j}+J'\sum_{[ij]}{\bf S}_{i}.{\bf S}_{j}+{\bf D}.\sum_{\langle ij\rangle}
{\bf S}_{i}\times{\bf S}_{j}-g\mu_{B}B\sum_{i}S_{i}^{z}
\label{e1}
\end{eqnarray}
where symbol $\langle ij\rangle$ and $[ij]$ implies the
nearest neighbor and next nearest neigbor sites in a honeycomb lattice, respectively.
$J$ is the nearest neighbor coupling constant between spins, while $J'$ is the next nearest neighbor
 coupling
constant between spins. The third term in Eq.(\ref{e1}) with ${\bf D}=(0,0,D)$ describes the
Dzyaloshinskii-Moriya interaction between nearest neighbor sites.
Also $g\simeq 2.2$ is the gyromagnetic constant and $\mu_B$ denotes the Bohr magneton.
 $B$ introduces the magnetic field strength.
Anisotropy due to DM interaction and Zeeman term violate
 SU(2) symmetry of the isotropic Heisenberg model Hamiltonian.
In order to obtain the bosonic representation of the model hamiltonian
two different bosonic operators are required.
Therefore spin operators are transformed to bosonic ones as
\begin{eqnarray}
S^{+}_{l,a}=a_{l}\;\;,\;\;S^{-}_{l,a}=a^{\dag}_{l}
\;\;,\;\;S^{z}_{l,a}=1/2-a^{\dag}_{l}a_{l}.\nonumber\\
S^{+}_{l,b}=b_{l}\;\;,\;\;S^{-}_{l,b}=b^{\dag}_{l}
\;\;,\;\;S^{z}_{l,b}=1/2-b^{\dag}_{l}b_{l}.
\label{e2}
\end{eqnarray}
$l$ denotes the unit cell index and $a,b$ label two sublattices.
$a^{\dag}_{l}(b^{\dag}_{l})$ creates a boson in unit cell with index $l$ on sublattice $a(b)$.
Exploiting above transformation, we have the following one particle bosonic hamiltonian
\begin{eqnarray}
 \mathcal{H}_{bil}&=&\frac{J}{2}\sum_{l,\Delta}(a^{\dag}_{l+\Delta}b_{l}+h.c.)
+\frac{J'}{2}\sum_{l,\Delta'}\Big(a^{\dag}_{l+\Delta'}a_{l}+b^{\dag}_{l+\Delta'}b_{l}\Big)\nonumber\\
&-&\frac{J}{2}\sum_{l}\Big(a^{\dag}_{l+\Delta}a_{l+\Delta}+a^{\dag}_{l}a_{l}+
b^{\dag}_{l+\Delta}b_{l+\Delta}+b^{\dag}_{l}b_{l}\Big)
\nonumber\\&-&\frac{J'}{2}\sum_{l}\Big(a^{\dag}_{l+\Delta'}a_{l+\Delta'}
+a^{\dag}_{l}a_{l}+b^{\dag}_{l+\Delta'}b_{l+\Delta'}
+b^{\dag}_{l}b_{l}\Big)+\frac{D}{2i}\sum_{l,\Delta}\Big(
a^{\dag}_{l+\Delta}b_{l}-b^{\dag}_{l}a_{l+\Delta}\Big)\nonumber\\&+
&g\mu_{B}B\sum_{l}\Big(a^{\dag}_{l}a_{l}+b^{\dag}_{l}b_{l}\Big)
\label{e3}
\end{eqnarray}
Lattice translational vectors connecting the nearest and next nearest unit cells are given by
\begin{eqnarray}
 {\bf R}_{\Delta_{1}}&=&{\bf i}\frac{\sqrt{3}}{2}+{\bf j}\frac{1}{2}\;\;,{\bf R}_{\Delta_{2}}={\bf i}\frac{\sqrt{3}}{2}-{\bf j}\frac{1}{2}\nonumber\\
{\bf R}_{\Delta_{3}}&=&-{\bf R}_{\Delta_{1}}\;\;,\;\;{\bf R}_{\Delta_{4}}=-{\bf R}_{\Delta_{2}}\;\;,\;\;
{\bf R}_{\Delta'_{1}}=-\frac{1}{2}{\bf j}\nonumber\\
{\bf R}_{\Delta'_{2}}&=&\frac{1}{2}{\bf j}
\label{e4}
\end{eqnarray}
The length of unit cell vector is set to one.
 In terms of Fourier
space transformation
of Bosonic operators, the bilinear part of model Hamiltonian is given by
\begin{eqnarray}
 \mathcal{H}_{bil}&=&\sum_{k}\mathcal{A}_{{\bf k}}(a^{\dag}_{{\bf k}}a_{k}+b^{\dag}_{{\bf k}}b_{{\bf k}})+
\sum_{{\bf k}}\Big[\mathcal{B}_{{\bf k}}a^{\dag}_{{\bf k}}b_{{\bf k}}+
\mathcal{B}^{*}_{{\bf k}}b^{\dag}_{{\bf k}}a_{{\bf k}}\Big].
\label{e5}
\end{eqnarray}
The coefficients in the above equation are
\begin{eqnarray}
\mathcal{A}_{{\bf k}}&=&g\mu_{B}B+\phi({\bf k})-\frac{3}{2}J-3J'\;\;,\;\;
\mathcal{B}_{{\bf k}}=\phi'({\bf k})+\phi"({\bf k}),\nonumber\\
\phi({\bf k})&=&J'\Big(cos(\frac{\sqrt{3}}{2}k_{x}+\frac{k_{y}}{2})+cos(
\frac{\sqrt{3}}{2}k_{x}-\frac{k_{y}}{2})+cos(k_{y})\Big),\nonumber\\
\phi'({\bf k})&=&\frac{J}{2}\Big(1+2e^{-i(\frac{\sqrt{3}}{2}k_{x})}cos(\frac{k_{y}}{2})\Big),\nonumber\\
\phi"({\bf k})&=&\frac{D}{2i}\Big(2e^{-i(\frac{\sqrt{3}}{2}k_{x})}cos(k_{y}/2)-1\Big).
\label{e5.1}
\end{eqnarray}
The
 wave vectors ${\bf k}$ are considered in the first Brillouin zone of the honeycomb lattice.
 Also the quartic part of the model Hamiltonian is obtained as follows
\begin{eqnarray}
 \mathcal{H}_{quartic}=\sum_{k,k',q}(\phi_{{\bf q}}+\phi'_{{\bf q}})a^{\dag}_{
{\bf k}+{\bf q}}b^{\dag}_{{\bf k'}-{\bf q}}b_{{\bf k'}}a_{{\bf k}} \;.
\label{e6}
\end{eqnarray}
In order to reproduce the SU(2) spin algebra, bosonic particles
 must also obey the local hard-core constraint
, i.e only one boson can occupy a single site of lattice.
In terms of Fourier transformation of bosonic operators,
we can write this part of the Hamiltonian as
\begin{eqnarray}
 \mathcal{H}_{int}=\mathcal{U}\sum_{k,k',q}(a^{\dag}_{
k+q}a^{\dag}_{k'-q }a_{k'}a_{k}+b^{\dag}_{
k+q}b^{\dag}_{k'-q }b_{k'}b_{k}).
\label{e7}
\end{eqnarray}
The effect of hard core repulsion part ($\mathcal U \rightarrow \infty$)
of the interacting Hamiltonian in Eq.(\ref{e4}) is dominant
compared with the quartic term in Eq.(\ref{e6}).
Using a unitary transformation as
\begin{eqnarray}
 \alpha_{{\bf k}}=v_{{\bf k}}a_{{\bf k}}+u_{{\bf k}}b_{{\bf k}}\;\;,\;\;
\beta_{{\bf k}}=-u^{*}_{{\bf k}}a_{{\bf k}}+v_{{\bf k}}b_{{\bf k}},
\label{e77777}
\end{eqnarray}
the bilinear part of the Hamiltonian is diagonalized as
\begin{eqnarray}
 \mathcal{H}_{bil}&=&\sum_{k}(\omega_{\alpha}({\bf k})\alpha^{\dag}_{{\bf k}}\alpha_{{\bf k}}
+\omega_{\beta}({\bf k})\beta^{\dag}_{{\bf k}}\beta_{{\bf k}})\nonumber\\
\omega_{\alpha}({\bf k})&=&\mathcal{A}_{{\bf k}}+|\mathcal{B}_{{\bf k}}|\;\;,\;\;
\omega_{\beta}({\bf k})=\mathcal{A}_{{\bf k}}-|\mathcal{B}_{{\bf k}}|.
\label{e88888}
\end{eqnarray}
The Bolgoliubov coefficients $u,v$ are given by
\begin{eqnarray}
 u_{k}=\sqrt{\frac{\phi'({\bf k})+\phi"
({\bf k})}{2(\phi'({\bf k})+\phi"
({\bf k}))}}\;\;,\;\;v_{k}=\frac{1}{\sqrt{2}}
\label{e99999}
\end{eqnarray}
The details of derivations of Bolgoliubov coefficients $u,v$ in Eq.(\ref{e99999}) and energy spectrums in Eq.(\ref{e88888}) are given in Appendix.
According to bilinear part of
 bosonic model Hamiltonian in Eq.(\ref{e5}),
Fourier transformation of the noninteracting
 Green's function matrix elements at finite temperature ($T$) are written in the following form
\begin{eqnarray}
 G^{(0)}_{aa}({\bf k},i\omega_{n})&=&-\int_{0}^{1/(k_{B}T)}d\tau e^{i\omega_{n}\tau}\langle
T(a_{{\bf k}}(\tau)a^{\dag}_{{\bf k}}(0))\rangle=
\sum_{j=\alpha,\beta}\frac{1}{2}(\frac{1}{i\omega_{n}
-\omega_{j}({\bf k})})=G^{(0)}_{bb}({\bf k},i\omega_{n}),\nonumber\\
 G^{(0)}_{ab}({\bf k},i\omega_{n})&=&-\int_{0}^{1/(k_{B}T)}d\tau e^{i\omega_{n}\tau}\langle T(a_{{\bf k}}
(\tau)b^{\dag}_{{\bf k}}(0))\rangle=
u_{{\bf k}}v_{{\bf k}}(\frac{1}{i\omega_{n}-\omega_{\alpha}({\bf k})}-
\frac{1}{i\omega_{n}-\omega_{\beta}({\bf k})}),\nonumber\\
G^{(0)}_{ba}({\bf k},i\omega_{n})&=&-\int_{0}^{1/(k_{B}T)}d\tau e^{i\omega_{n}\tau}\langle
T(b_{{\bf k}}(\tau)a^{\dag}_{{\bf k}}(0))\rangle=
u^{*}_{{\bf k}}v_{{\bf k}}(\frac{1}{i\omega_{n}-\omega_{\alpha}({\bf k})}
-\frac{1}{i\omega_{n}-\omega_{\beta}({\bf k})}),
\end{eqnarray}
where $\omega_{n}=2n\pi k_{B}T$ denotes the bosonic
Matsubara's frequency. Since the hard core bosonic repulsion is on-site interaction between bosons,
it is predicted that only the diagonal elements of bosonic self-energy matrix gets non zero value.
Thus
we can write down a Dyson's equation for the interacting Green's
function matrix (${\bf G}({\bf k},i\omega_{n})$)  as
\begin{eqnarray}
 {\bf G}({\bf k},i\omega_{n})&=&{\bf G}^{(0)}({\bf k},i\omega_{n})+{\bf G}^{(0)}({\bf k},i\omega_{n}){\bf \Sigma}({\bf k},i\omega_{n}){\bf G}({\bf k},i\omega_{n}),\nonumber\\
{\bf G}^{(0)}({\bf k},i\omega_{n})&=&\left(
                              \begin{array}{cc}
                               G^{(0)}_{aa}({\bf k},i\omega_{n}) &  G^{(0)}_{ab}({\bf k},i\omega_{n}) \\
                              G^{(0)}_{ba}({\bf k},i\omega_{n}) & G^{(0)}_{bb}({\bf k},i\omega_{n}) \\
\end{array}
\right),\nonumber\\
{\bf \Sigma}({\bf k},i\omega_{n})&=&\left(
                              \begin{array}{cc}
                               \Sigma_{aa}({\bf k},i\omega_{n}) & 0 \\
                             0 & \Sigma_{bb}({\bf k},i\omega_{n}) \\
\end{array}
\right),
\label{e7.0}
\end{eqnarray}
where $\Sigma^{Ret}_{aa}({\bf k},\omega)=\Sigma^{Ret}_{bb}({\bf k},\omega)$ is normal
retarded self-energies diagonal element due to hard core repulsion between bosons.
After using Dyson's series for
the interacting Matsubara Green's function matrix in Eq.(\ref{e7.0})
\cite{mahan}, the low energy limit of single particle retarded Green's function is
\begin{equation}
G_{aa}^{sp}({\bf k},\omega)= G_{aa}({\bf k},i\omega_{n}\longrightarrow\omega+i0^{+})=
\frac{1}{2}\sum_{j=\alpha,\beta}\frac{Z_{{\bf k}}}{\omega-\Omega_{\alpha}({\bf k})+i0^{+}}.
\label{e12}
\end{equation}
The renormalized excitation spectrum and renormalized single particle weight are given by
\begin{eqnarray}
\Omega_{j=\alpha,\beta}({\bf k})&=&Z_{{\bf k}}(\omega_{j=\alpha,\beta}({\bf k})+\Sigma_{aa}({\bf k},0)),
\nonumber\\
Z_{{\bf k}}^{-1}&=&1-(\frac{\partial Re\Big(\Sigma^{Ret}_{aa}({\bf k},\omega)\Big)}{\partial \omega})
_{\omega=0},
\label{e13}
\end{eqnarray}
 Since the Hamiltonian
$\mathcal{H}_{int}$ in Eq.(\ref{e4}) is short ranged and $\mathcal{U}$ is
large, the Brueckner's approach (ladder diagram summation)
\cite{rezania2008,fetter,gorkov} can be employed for calculating bosonic self-energies in
 the low density limit of
bosonic gas and for low temperature. Firstly, the scattering
amplitude (t-matrix) $\Gamma(p_{1},p_{2};p_{3},p_{4})$ of hard core bosons is
introduced where $p_{i}\equiv({\bf p},(p_{0}))_{i}$.
The basic approximation made in the derivation of $\Gamma(K\equiv
p_{1}+p_{2})$ is that we have considered the diagonal elements of bosonic Green's function for \
making $\Gamma$. It is
 due to the strongly intrasite interaction between bosons.
According to the Feynman's rules
in momentum space at finite temperature and
after taking limit $\mathcal{U}\longrightarrow\infty$, the scattering
amplitude can be written as the following form
\begin{eqnarray}
\Gamma_{aaaa}({\bf K},i\omega_{n})&=&-\Big(\frac{k_{B}T}
{N}\sum_{{\bf Q},m}
  G^{(0)}_{aa}({\bf Q},iQ_{m})
G^{(0)}_{aa}({\bf K}-{\bf Q},i\omega_{n}-iQ_{m}),
\Big)^{-1}.
 \label{e178}
\end{eqnarray}
where $N$ is the number of unit cells and wave vector ${\bf Q}$ belongs to the
 first Brillouin zone of honeycomb lattice.
We can perform the summation over Matsubara frequencies $Q_{m}=2m\pi k_{B}T$
in the Eq.(\ref{e178}) according to Feynman
rules\cite{mahan} and the final result for scattering amplitude is obtained by
\begin{eqnarray}
\Gamma_{aaaa}({\bf K},i\omega_{n})&=&-\Big(\frac{1}
{N}
\sum_{{\bf Q}} \frac{1}{4}[
\sum_{j,j'=\alpha,\beta}(\frac{n_{B}(\omega_{j}({\bf Q}))-n_{B}
(-\omega_{j'}({\bf K}-{\bf Q}))}
{i\omega_{n}-\omega_{j}({\bf Q})-\omega_{j'}({\bf K}-{\bf Q})}
\Big)^{-1},
 \label{e178.5}
\end{eqnarray}
where $n_{B}(x)=1/(e^{x/(k_{B}T)}-1)$ denotes Bose-Einstein distribution function.
The hard core self-energy is obtained by using the vertex-function
obtained in Eq.(\ref{e178}).
\begin{equation}
\Sigma_{aa}(\textbf{k},i\omega_{n})=-\frac{k_{B}T}{N}\sum_{p_{m},{\bf p}}
\Gamma_{aaaa}(p,k;p,k)G^{(0)}_{aa}({\bf p},ip_{m})-\frac{k_{B}T}{N}
\sum_{p_{m},{\bf p}}\Gamma_{aaaa}(p,k;k,p)
G^{(0)}_{aa}({\bf p},ip_{m}). \hspace{3mm}
\label{e190}
\end{equation}
The hard core self-energy is obtained by taking integration
over internal energy ($p_{m}$)
\begin{eqnarray}
\Sigma_{aa}({\bf k},i\omega_{n})=\frac{1}{N}\sum_{{\bf p}}\Big(n_{B}(\omega_{\alpha}({\bf p}))
\Gamma({\bf p}+{\bf k},\omega_{\alpha}({\bf p})+i\omega_{n})+
n_{B}(\omega_{\beta}({\bf p}))\Gamma({\bf p}+{\bf k},\omega_{\beta}({\bf p})+i\omega_{n})\Big).
 \label{e211}
\end{eqnarray}
We can simply obtain the retarded self-energy by analytic
continuation ($i\omega_{n}\longrightarrow\omega+i0^{+}$) of
Eq.(\ref{e211}). The contribution of $\mathcal{H}_{quartic}$ on the
final results is very small because it is composed of quartic terms
in the bosonic operators. It is therefore treated in mean-field
approximation. This is equivalent to take only one-loop diagrams in
to account (first order in $J$). On the mean field level we have
$O_{1}O_{2}=\langle O_{1}\rangle O_{2}+ \langle O_{2}\rangle
O_{1}-\langle O_{1}\rangle\langle O_{2}\rangle$ where each $O_{1}$
and $ O_{2}$ is a pair of boson operators. We can write for each
pair of operators,
\begin{eqnarray}
\langle a^{\dag}_{{\bf k}}a_{{\bf k}'}\rangle&=&-\delta_{{\bf k},{\bf k}'}\frac{1}{\beta N}\sum_{n}
G^{(0)}_{aa}({\bf k},i\omega_{n})=\frac{1}{2N}\Big(n_{B}(\omega_{\alpha}({\bf k}))+n_{B}(
\omega_{\beta}({\bf k}))\Big),\nonumber\\
\langle a^{\dag}_{{\bf k}}b_{{\bf k}'}\rangle&=&-\delta_{{\bf k},{\bf k}'}
\frac{1}{\beta N}\sum_{n}
G^{(0)}_{ab}({\bf k},i\omega_{n})=\frac{u_{{\bf k}}v_{{\bf k}}}{2N}\Big(n_{B}(\omega_{\alpha}
({\bf k}))+n_{B}(
\omega_{\beta}({\bf k}))\Big).
\label{e211.1}
\end{eqnarray}
Thus, the effect of $H_{quartic}$ is to renormalize the coefficients of bilinear part of
the Hamiltonian according to the following relations
\begin{eqnarray}
\mathcal{A}_{{\bf k}}&\rightarrow&\mathcal{A}_{{\bf k}}+
(\phi_{{\bf q}=0}+\phi'_{{\bf q}=0})\frac{1}{N}\sum_{{\bf k}}\Big(n_{B}(\omega_{\alpha}({\bf k}))+n_{B}(
\omega_{\beta}({\bf k}))\Big),\nonumber\\
\mathcal{B}_{{\bf k}}&\rightarrow&\mathcal{B}_{{\bf k}}+
(\phi_{{\bf q}=0}+\phi'_{{\bf q}=0})\frac{1}{2N}\sum_{{\bf k}} u_{{\bf k}}
v_{{\bf k}}\Big(n_{B}(\omega_{\alpha}({\bf k}))+n_{B}(
\omega_{\beta}({\bf k}))\Big).
\label{e211.10}
\end{eqnarray}
The renormalized coefficients (Eq.(\ref{e211.10})) will be considered to calculate
the self-energy which are independent of energy (nonretarded in time representation).
\section{Magnetic specific heat and longitudinal magnetization}
In order to obtain magnetic specific heat, we will obtain the
internal energy in terms of one particle Green's function by the
equation of motion. Then the specific heat is obtained from the
temperature derivative of internal energy, $C_{V}=\frac{\partial
E}{\partial T}$. The effective model Hamiltonian including the hard core repulsion between
bosons is given by
\begin{eqnarray}
 \mathcal{H}_{bil}=\sum_{{\bf k}}(\Omega_{\alpha}({\bf k})\alpha^{\dag}_{{\bf k}}\alpha_{{\bf k}}
+\Omega_{\beta}({\bf k})\beta^{\dag}_{{\bf k}}\beta_{{\bf k}}).
\end{eqnarray}
Using Bose Einstein distribution function, thermal quantum average
of the above model Hamiltonian gives the following specific heat as
\begin{eqnarray}
C_{v}=\frac{d}{dT}\Big(\sum_{k}\frac
{\Omega_{\alpha}({\bf k})}{e^{\Omega_{\alpha}({\bf k})/k_{B}T}-1}
+\frac
{\Omega_{\beta}({\bf k})}{e^{\Omega_{\beta}({\bf k})/k_{B}T}-1}\Big)
\label{a131}
\end{eqnarray}
The magnetization along magnetic field ($M$) can be expressed in terms of
density of hard core bosons
\begin{eqnarray}
M=1-\frac{1}{N}\sum_{{\bf k}}\langle a^{\dag}_{{\bf k}}a_{{\bf k}}\rangle-
\frac{1}{N}\sum_{{\bf k}}\langle b^{\dag}_{{\bf k}}b_{{\bf k}}\rangle
\label{e131}
\end{eqnarray}
$\langle a^{\dag}_{k}a_{k}\rangle$ and $\langle b^{\dag}_{k}b_{k}\rangle$  can be related
to the one particle interacting
 Green's function. After using interacting Green's function in Eq.(\ref{e12}) and performing
some algebra calculations, we arrive the following expression for longitudinal magnetization
\begin{eqnarray}
M=1-\frac{1}{N}\sum_{{\bf k}}\Big(n_{B}(\Omega_{\alpha}({\bf k}))+n_{B}(
\Omega_{\beta}({\bf k}))\Big),
\label{e1333}
\end{eqnarray}
where $N$ is the number of unit cells of honeycomb lattice and
$n_{B}$ is the bosonic distribution function. The relations for
$\Omega_{\alpha}({\bf k})$ and
 $\Omega_{\beta}({\bf k})$
have been given in Eq.(\ref{e13}).

\section{Results and discussions}
In this article we have studied the effects of DM interaction strength, next nearest neighbor coupling exchange constant and magnetic field
on the thermodynamic properties
 of the spin 1/2 Heisenberg model on honeycomb lattice in the field induced spin-polarized phase.
Specially,
we mostly concentrate on the behavior of specific heat and longitudinal magnetization
 versus temperature, magnetic field and anisotropic DM interaction.
The Dzyaloshinskii-Moriya interaction breaks the SU(2) symmetry of the model
and changes the properties of the model which is discussed
in this section. Thermodynamic properties of anisotropic Heisenberg model on honeycomb lattice have been discussed
using excitation spectrum of hard core bosonic gas.
The original spin model has been represented by a bosonic model in
the presence of hard core repulsion to avoid double occupation of
bosons at each lattice site which preserves the SU(2) algebra of the
spin model Hamiltonian. In the limit of $B/J\longrightarrow\infty$, the ground
state is a field induced spin-polarized state and a finite energy
gap exists to the lowest excited state.
 The decrease in magnetic field lowers the excitation gap
which eventually vanishes at the critical magnetic field ($B_c$).
 We
have implemented the Green's function approach to obtain the effect
of interaction on the diagonal part of the bosonic Hamiltonian using
Brueckner's formalism above threshold field $B_{c}$ where the density of bosonic
gas is small.

The single particle excitation should be found from a self-consistent
solution of Eqs.(\ref{e13},\ref{e178.5},\ref{e211},\ref{e211.10})
with the substitutions
$u_{k}\longrightarrow\sqrt{Z_{k}}U_{k}$, $v_{k}\longrightarrow\sqrt{Z_{k}}
V_{k}$, $\omega_{k}\longrightarrow\Omega_{k}$ in the corresponding equations.
The process is started with an initial guess for $Z_{k},\Sigma_{aa}(k,0)$ and
 by using Eq.(\ref{e13}) we find corrected excitation energy.
This is repeated until convergence is reached.
Using the final values for excitation spectrum
we can calculate specific heat and longitudinal magnetization
 by Eqs.(\ref{a131},\ref{e1333}), respectively. We discuss the numerical results for
thermodynamic properties in the field induced spin polarized regime where
 energy
spectrum of spin model hamiltonian includes a finite energy gap
between ground state and first excited state. Therefore as long as
excitation spectrum $\Omega_{{\bf Q}_{0}=(0,4\pi/3)}$ has non zero values, the system
preserves its gapped spin polarized phase.

In Fig.1(a) we have plotted the
energy gap ($\Delta$) versus magnetic field $g\mu_{B}B/J$
for different values of next nearest neighbor coupling exchange constant
 $J'/J$ for $D/J=0.2$ by setting $k_{B}T/J=0.05$.
It is obvious from Fig.1(a) that the energy gap
vanishes as the magnetic field approaches the critical value for
$g\mu_{B}B_{c}/J$. For all values of $J'/J$, the gap vanishes at the critical point
 $g\mu_{B}B_{c}/J$
where the transition from gapped spin liquid phase to the gapless
one occurs. According to Fig.1(a), the critical field increases with
$J'/J$. For magnetic fields above critical field $g\mu_{B}B_{c}/J$,
energy gap
 exists to the lowest excited state which is called spinon spectrum.
Decreasing the magnetic field leads to vanish the energy gap and a gapless magnetic ordering
state
develops for magnetic fields below $g\mu_{B}B_{c}/J$ for each $J'/J$.
According to this figure the magnetic field region
where excitation spectrum becomes gapless grows with anisotropy parameter.
In other words the field induced spin-polarized phase sets up in
lower magnetic field with decrease of $J'/J$.

The effect of DM interaction strength, $D$, on critical magnetic field has been studied
in Fig.1(b). In this figure,
energy gap ($\Delta$) versus magnetic field $g\mu_{B}B/J$
for different values of Dzyaloshinskii-Moriya interaction strength
 $D/J$ for $J'/J=0.2$ by setting $k_{B}T/J=0.05$ has been plotted.
Fig.1(b) shows that the energy gap
vanishes as the magnetic field approaches the critical value for
$g\mu_{B}B_{c}/J$. For all values of $D/J$, the gap vanishes at the critical point
 $g\mu_{B}B_{c}/J$
where the transition from gapped spin liquid phase to the gapless
one occurs. Moreover the critical field tends to higher value with
increase of $D/J$ according to Fig.1(b).

The temperature behavior of magnetic specific heat of localized electrons on honeycomb lattice
  for various magnetic fields
has been plotted in Fig.(2) by setting $D/J=J'/J=0.2$.  Since our approach is based on the high magnetic
 field limit the value of
magnetic field is restricted to $B>B_{c}$ where $B_{c}$ is the critical field at
finite temperature. Each curve shows an exponential decay at low temperature which manifests
the presence of a finite-energy gap. Larger values of $g\mu_{B}B/J$ show more rapid decay corresponding
to larger energy gap.
This increasing behavior of specific heat
 is in agreement with experimental
measurements\cite{hong}. In this study the temperature behavior of
specific heat has been measured for magnetic fields above and below
threshold field. Fig.(2) indicates the increase of magnetic
field leads to decrease of specific heat for all temperature region.
 It can be understood from the fact the energy gap width grows with increase of
magnetic field $g\mu_{B}B/J$.
In other hand the low temperature limit of specific heat is proportional to
$1/Te^{-\Delta/T}$ and therefore the decrease of specific heat with magnetic field
 can be justified.
 Since hard core bosons behave as classical objects at high temperatures, specific
heat gets the constant value in this temperature region for all
magnetic fields as shown in Fig.(2). Specific heat becomes constant
in temperature region above characteristic temperature 2.5 for
normalized magnetic field $g\mu_{B}B/J=5.5$. The increase of
magnetic field leads to increase of this characteristic temperature
according to Fig.(2). Similar behaviors of specific heat of
localized electrons on honeycomb lattice have been obtained by
numerical results. In a numerical calculations based on high
temperature series expansions\cite{singh} the specific heat and
susceptibility of honeycomb lattice Heisenberg model have been
studied. An increasing behavior for specific heat at low
temperatures has been obtained in this work. Such results for
specific heat is in agreement with our results for specific heat of
localized electrons on honeycomb lattice. Also in the other
numerical work, temperature dependence of magnetic structure factors
and specific heat of Heisenberg model on honeycomb structure has
been investigated using numerical exact diagonalization
method\cite{yamaji}. Moreover the staggered magnetization and
specific of
 Heisenberg model Hamiltonian on honeycomb lattice has been studied by exploiting numerical quantum monte carlo method\cite{huang}.
The general behaviors of temperature dependence of specific heat is in agreement with our results.
The magnetic properties of the two-dimensional S = 1/2 quantum antiferromagnetic Heisenberg model
on a honeycomb lattice are studied by means of a continuous Euclidean time Quantum-Monte-Carlo algorithm\cite{low}.

The effect of next nearest neighbor coupling exchange constant
 $J'/J$ on the temperature behavior of specific heat at fixed magnetic field
$g\mu_{B}B/J=6.0$ for $D/J=0.2$ is shown in Fig.3(a).
Here $J'$ has a direct influence
on the energy gap and hence on the bosonic density as shown in Fig.3(a).
The increase of $J'/J$ raises the energy
 gap which gives lower specific heat at a given normalized
temperature where specific heat behaves as $1/Te^{-E_{g}/T}$.
Furthermore Fig.3(a) implies specific heat reaches a constant value
for temperatures above a characteristic temperature. According to
Fig.3(a) this characteristic temperature goes to lower value with
$J'$. This can be understood from this fact that increase of $J'$
leads to decrease of energy gap and thus transition of bosons from
ground state to excited state are performed at lower temperatures.
Consequently classical behavior of bosons begins at higher
temperatures with decrease of next nearest neighbor coupling
exchange constant $J'$.

In Fig.3(b), we plot specific heat versus normalized temperature for different values of
DM interaction strength
, namely $D/J=0.2,0.6,0.8,1.5$ for $J'/J=0.2$ at fixed magnetic field
$g\mu_{B}B/J=6.0$. This plot indicates that specific heat increases with temperature for each value
of $D/J$ up to a characteristic temperature. Upon increasing temperature above
characteristic one, specific heat gets a constant value.
The characteristic temperature tends to lower amounts with $D/J$ as shown in Fig.3(b).
This can be justified from the fact that energy gap decreases with $D$ which consequently reduces
this characteristic temperature.

In Fig.4(a), we plot specific heat versus normalized magnetic field
$g\mu_{B}B/J$ for different values of next nearest neighbor coupling exchange constant
 at fixed normalized
temperature $k_{B}T/J=0.03$ for $D/J=0.2$. This plot indicates a monotonic
decrease of specific heat for all values of $J'/J$ on the whole range
of magnetic field. It can be understood from the fact that energy gap
grows with magnetic field and consequently thus the Bosonic density
reduces. This fact leads to the decrease of the specific heat. Also Fig.4(a)
shows the specific heat goes to zero at magnetic fields above 5.9 for
all values of $J'/J$.
For each anisotropy parameter $\nu$, we have a different critical field,
so that the hard core Bosonic representation works fine for magnetic
fields above the critical field.
According to Fig.4(a),
specific heat increases with $J'/J$ for fixed normalized magnetic
 field. This behavior arises from this point that energy gap reduces with $J'$
which leads to increase specific heat.

Similar behavior has been obtained for
 DM interaction effects on magnetic field dependence of
specific heat. We have plotted specific heat of Heisenberg model Hamiltonian on honeycomb lattice
 as a function of normalized magnetic field for different values of
 $D/J$ for $J'/J=0.2$ by setting $k_{B}T/J=0.03$ in Fig.4(b).
A decreasing behavior for magnetic field dependence of specific heat is clearly observed
for each $D/J$ due to increase of energy gap with magnetic field.
Moreover specific heat enhances with increase of DM interaction strength.
This arises from this fact that energy gap reduces with $D/J$ which leads to increase of
specific heat.

We have also studied the behavior of longitudinal magnetization along perpendicular
to the plane for localized electrons
on two dimensional honeycomb lattice described by Heisenberg model Hamiltonian.
Fig.5(a) shows longitudinal magnetization ($M$)
 as a function of magnetic field $g\mu_{B}B/J$ for different values of $J'/J$.
For each value of $J'/J$,
 $M$ increases with magnetic field. The population of bosons decreases with magnetic field
and consequently magnetization increases based on Eq.(\ref{e131}).
Upon increasing normalized magnetic field ($g\mu_{B}B/J$) above 9.0
the magnetization reaches its saturate value. All curves fall on each other in magnetic field
region above 9.0 according Fig.5(a).

We have also studied the effect of DM interaction strength on the magnetic field dependence
 of
longitudinal magnetization of the system. In Fig.5(b), we plot $M$
versus normalized magnetic field for different values of DM integration
strength, namely $D/J=0.0,0.5,1.0,1.5,2.0$ for $J'/J=0.2$ at fixed
normalized temperature $k_{B}T/J=0.1$. This plot indicates the
increase of magnetic field raises
 $M$ for all values of $D/J$ . This fact can be understood that magnetic field
causes to decrease of bosonic density and therefore magnetization
grows with temperature. Upon more increasing normalized
 magnetic field above 8.5, $M$ gets its saturate value for all values of $D/J$.
At higher values of magnetic field above 8.0, the magnetization is independent of
Dzyaloshinskii-Moreover interaction strength and all curves fall on each other in this magnetic
field region as shown in Fig.5(b).

Finally we have studied the dependence of longitudinal magnetization on DM interaction strength
for various $J'/J$ for $g\mu_{B}B/J=6.0$ in Fig.(6).
Magnetization shows
 no considerable dependence on $D$ in the region $D/J<1.0$ for each value of next nearest neighbor
coupling exchange constant $J'/J$. Upon more increasing $D/J$ above
1.0, the magnetization reduces. This behavior can be understood from
this fact that Dzyaloshinskii-Moriya interaction leads to coupling
between the transverse components of spin operators which
consequently decreases the magnetic ordering along perpendicular to
the plane. However the slope of the reduction increases with $J'/J$
based on Fig.(6).
 Moreover the magnetization decreases with next nearest neighbor
coupling exchange constant $J'/J$ at fixed $D/J$.

\newpage
\begin{figure}
\includegraphics[width=8cm]{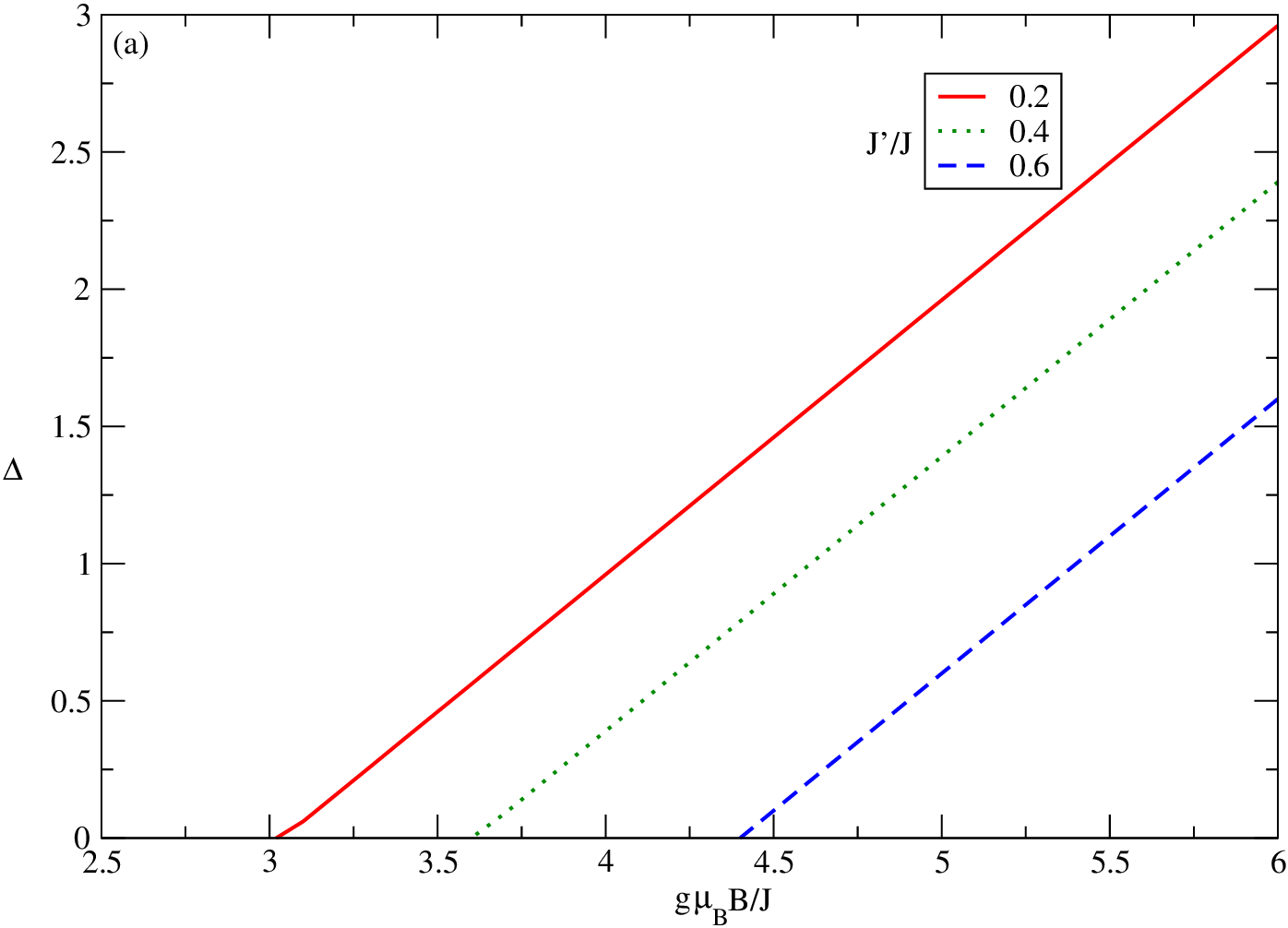}
\includegraphics[width=8cm]{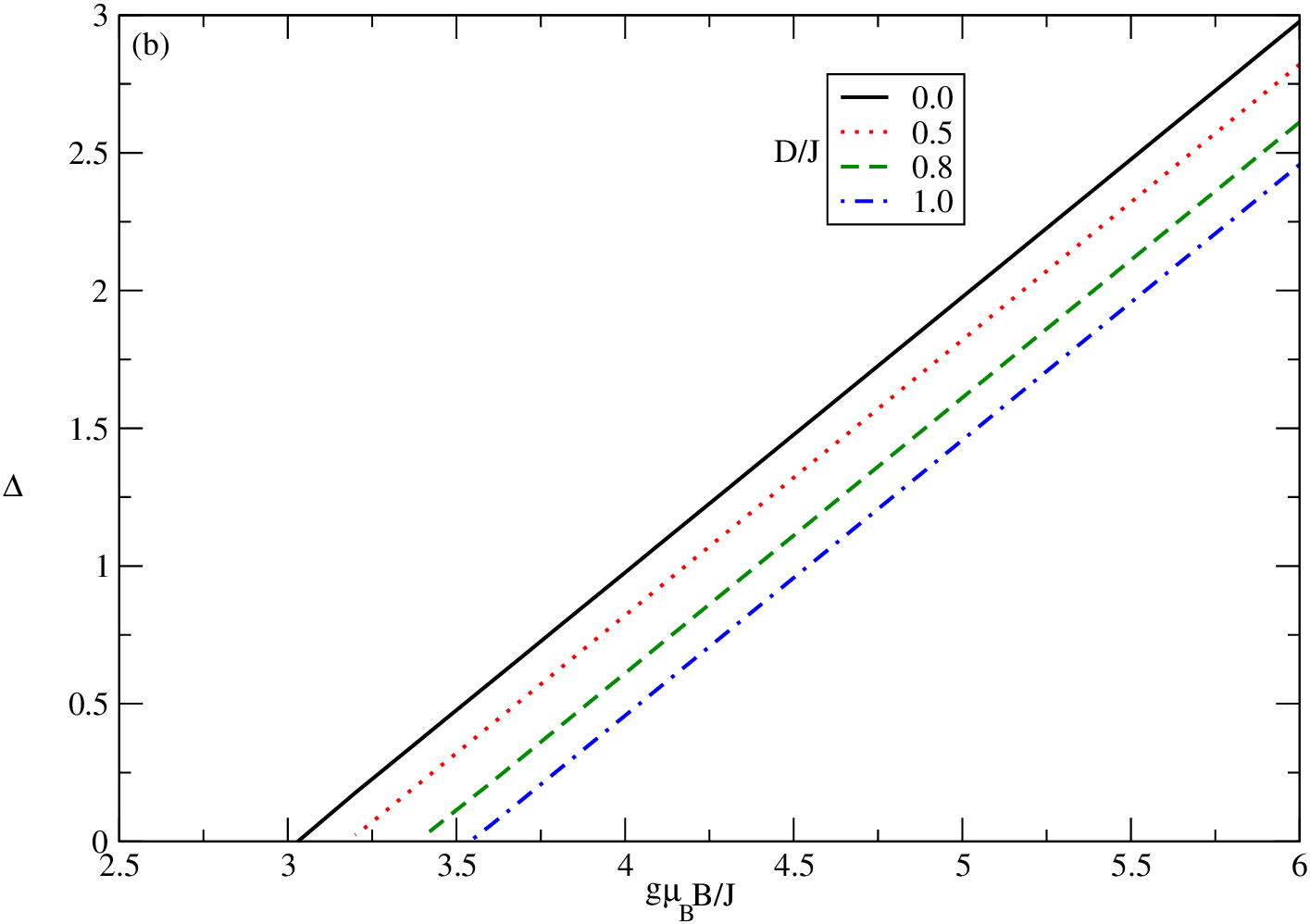}
\caption{(a) Energy gap ($\Delta$) versus
magnetic field ($g\mu_{B}B/J$) for $D/J=0.2$ and different values
$J'/J$ by setting $kT/J=0.05$.
(b)Energy gap ($\Delta$) versus
magnetic field ($g\mu_{B}B/J$) for $J'/J=0.2$ and different values
$D/J$ by setting $kT/J=0.05$.
The change in the critical magnetic field (where the gap vanishes)
for various anisotropies is remarkable.}
\end{figure}
\begin{figure}
\includegraphics[width=8cm]{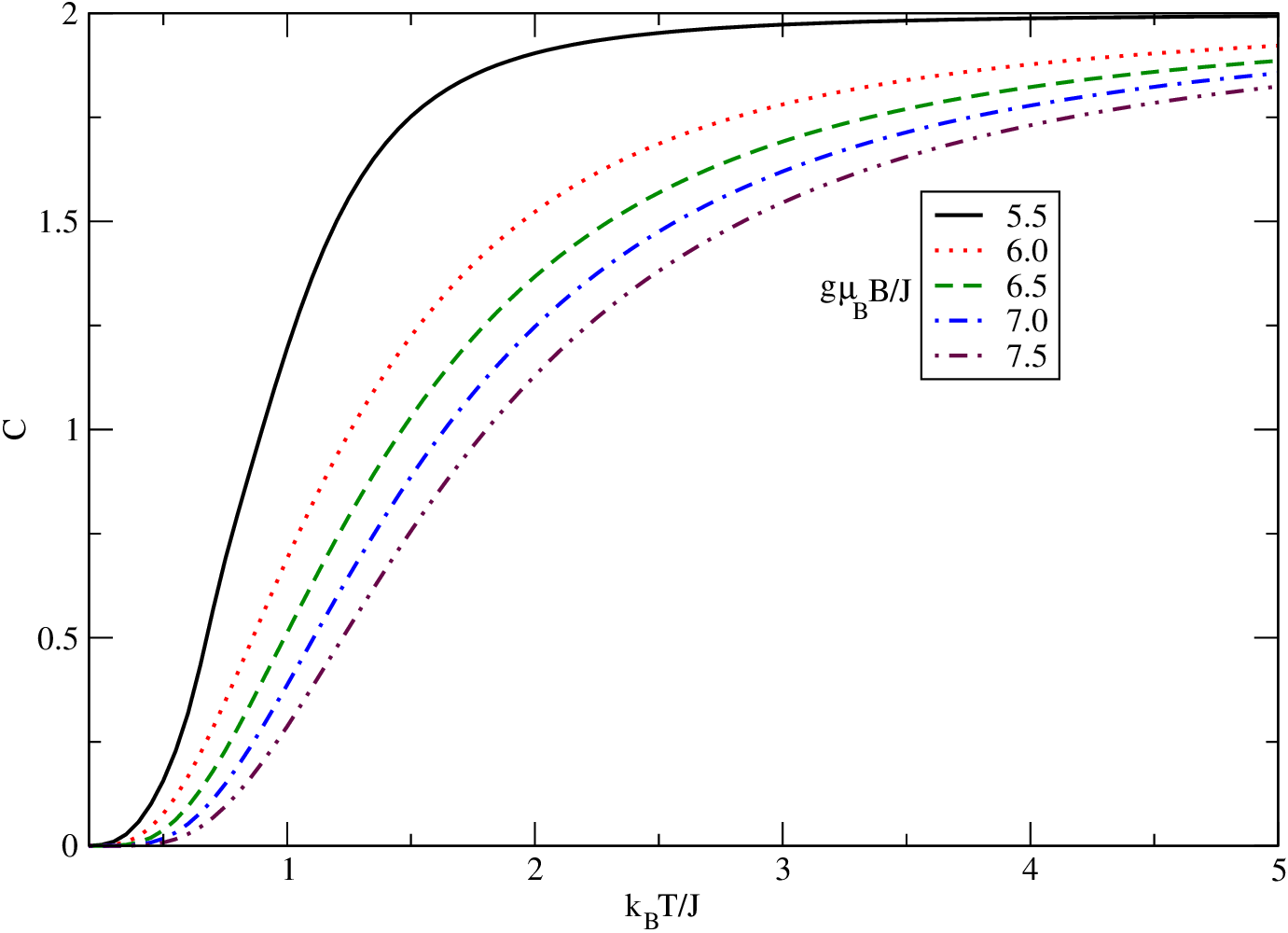}
\caption{Magnetic specific heat as a function of $k_{B}T/J$ for different values of
magnetic fields for fixed Dzyaloshinskii-Moriya interaction parameter $D/J=0.2$ and $J'/J=0.2$. }
\end{figure}
\begin{figure}
\includegraphics[width=8cm]{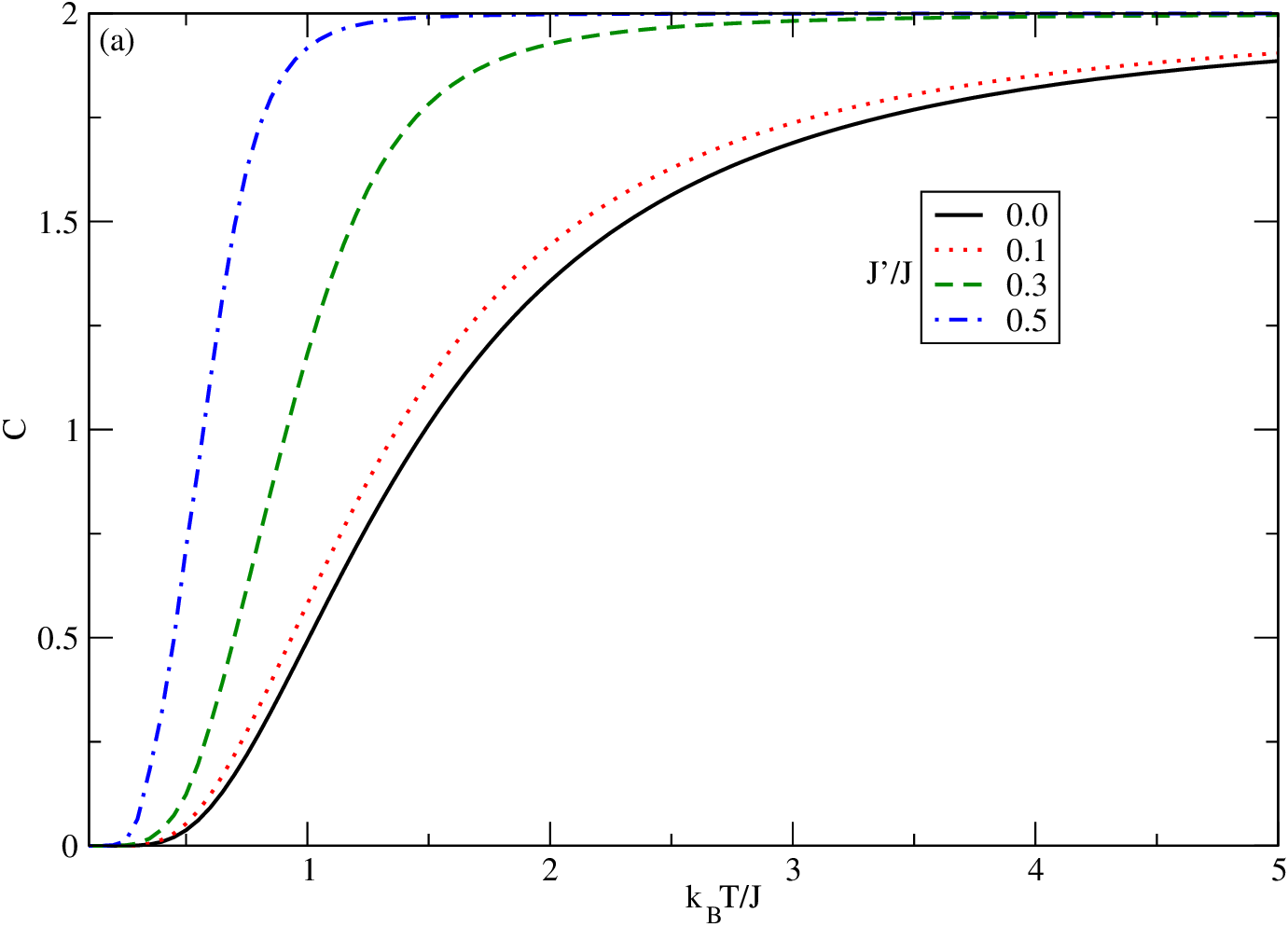}
\includegraphics[width=8cm]{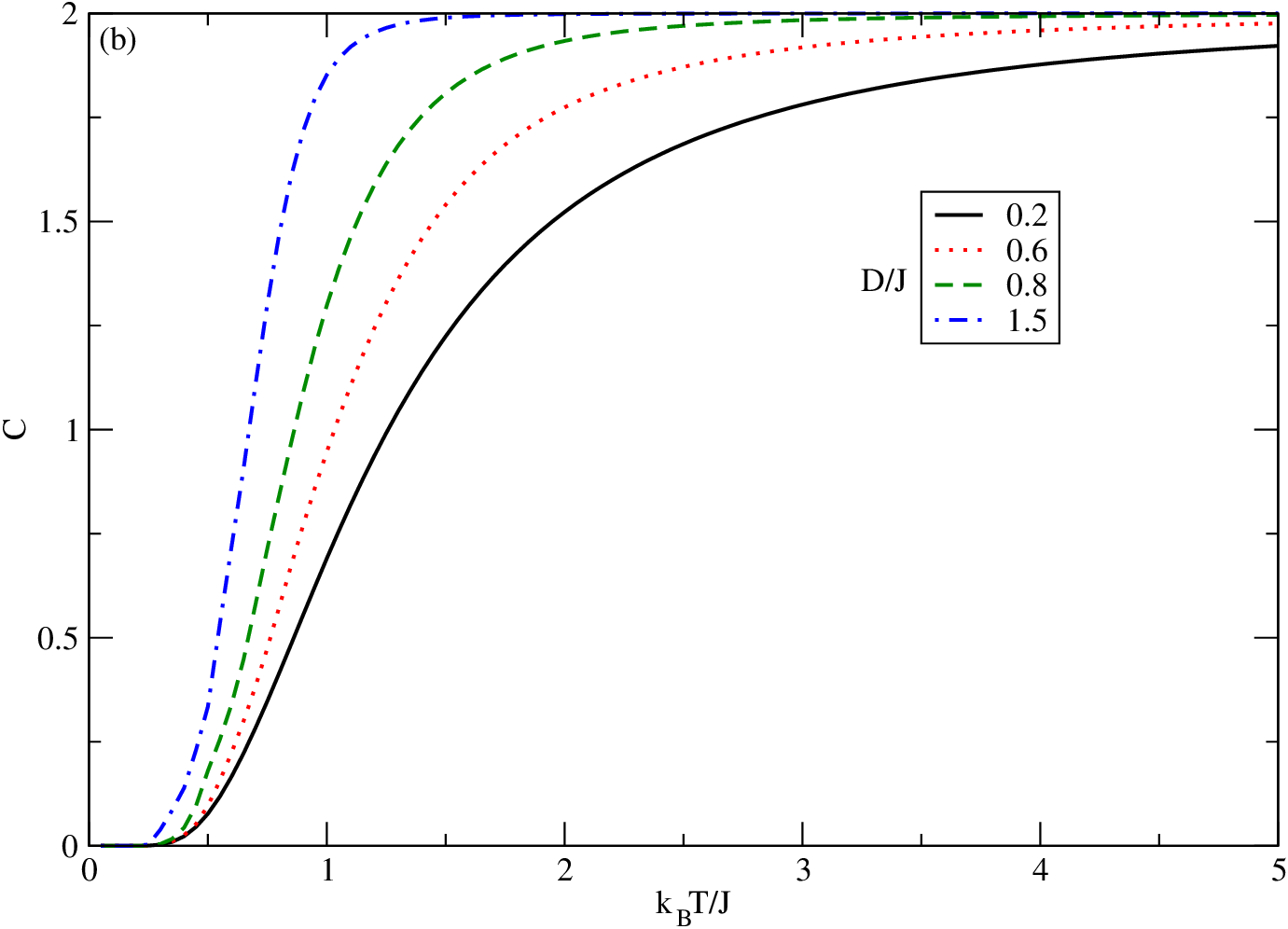}
\caption{(a) Magnetic specific heat as a function of $k_{B}T/J$ for different values of
next nearest neighbor coupling exchange constant for fixed
 Dzyaloshinskii-Moriya interaction parameter $D/J=0.2$. Also renormalized
 magnetic field
 is fixed at $g\mu_{B}B/J=6.0$ above threshold magnetic field.
(b) Magnetic specific heat as a function of $k_{B}T/J$ for different values of
 Dzyaloshinskii-Moriya interaction parameter for fixed
next nearest neighbor coupling exchange constant $J'/J=0.2$ and
$g\mu_{B}B/J=6.0$.}
\end{figure}
\begin{figure}
\includegraphics[width=8cm]{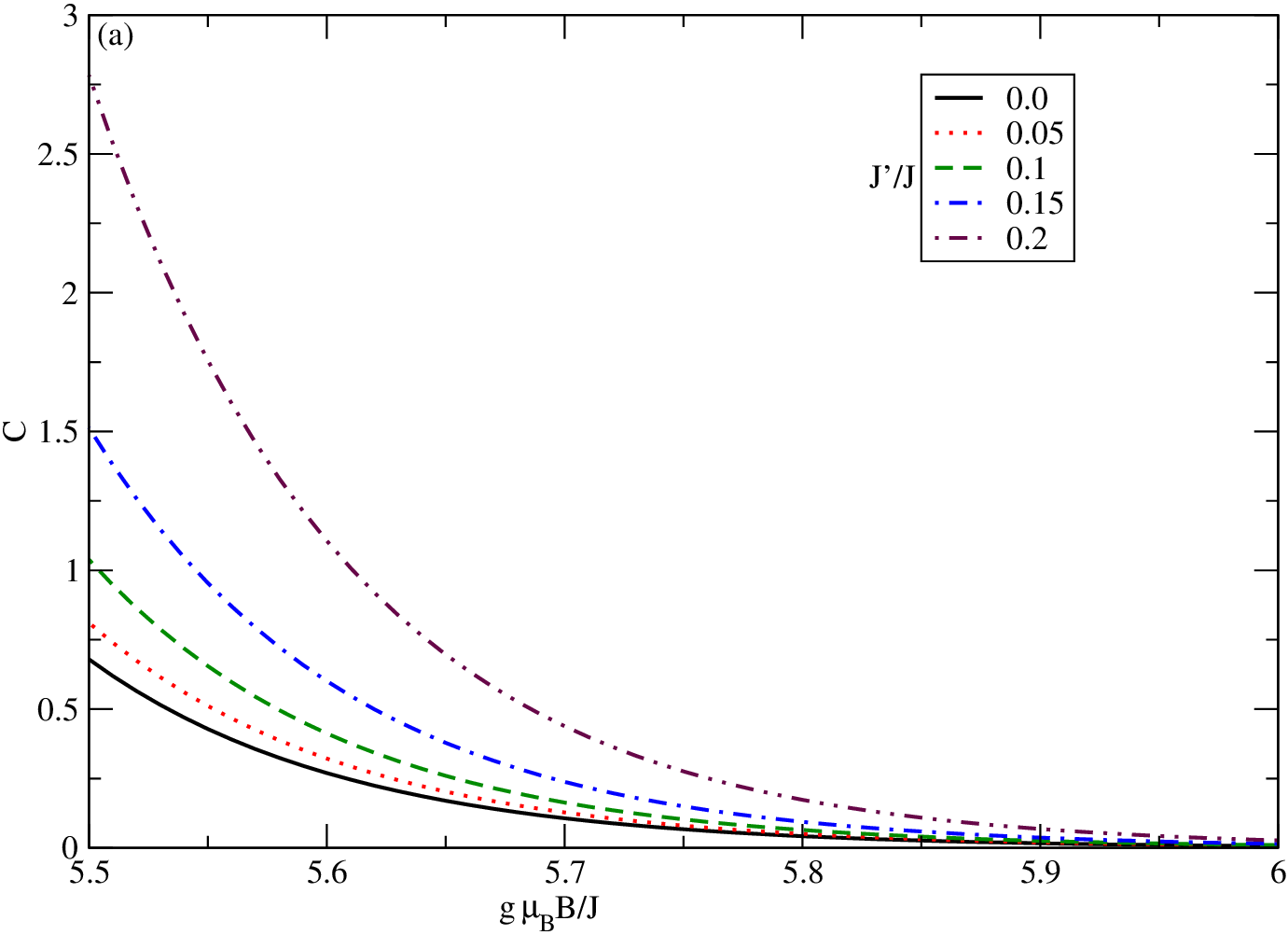}
\includegraphics[width=8cm]{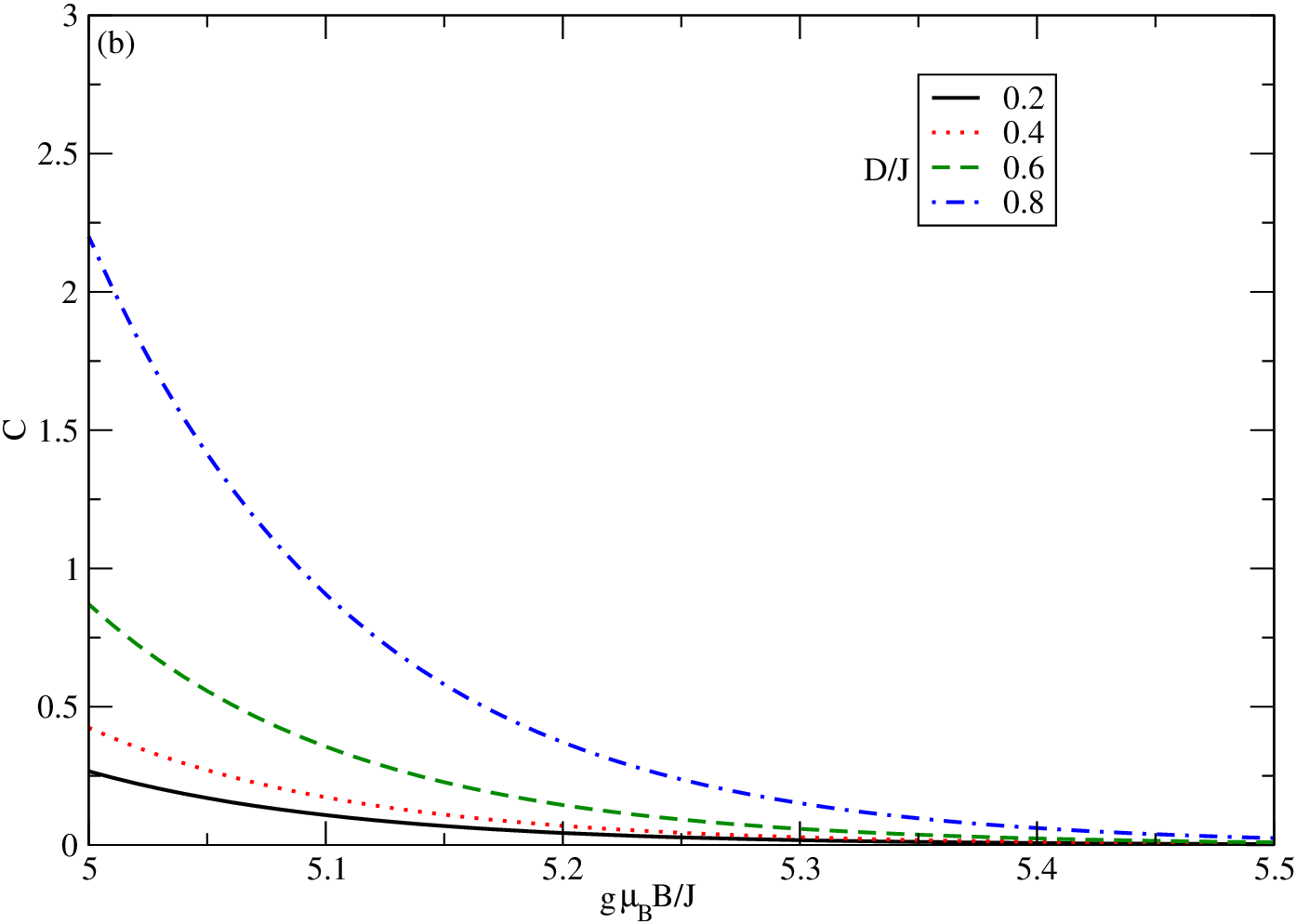}
\caption{(a) Magnetic specific heat as a function of normalized magnetic field $g\mu_{B}B/J$
for different values of
 next nearest neighbor coupling exchange constant $J'/J$ for $D/J=0.2$ and $k_{B}T/J=0.1$.
(b) Magnetic specific heat as a function of normalized magnetic field $g\mu_{B}B/J$
for different values of
 Dzyaloshinskii-Moriya interaction parameter $D/J$ for $J'/J=0.2$ and $k_{B}T/J=0.1$.}
\end{figure}
\begin{figure}
\includegraphics[width=8cm]{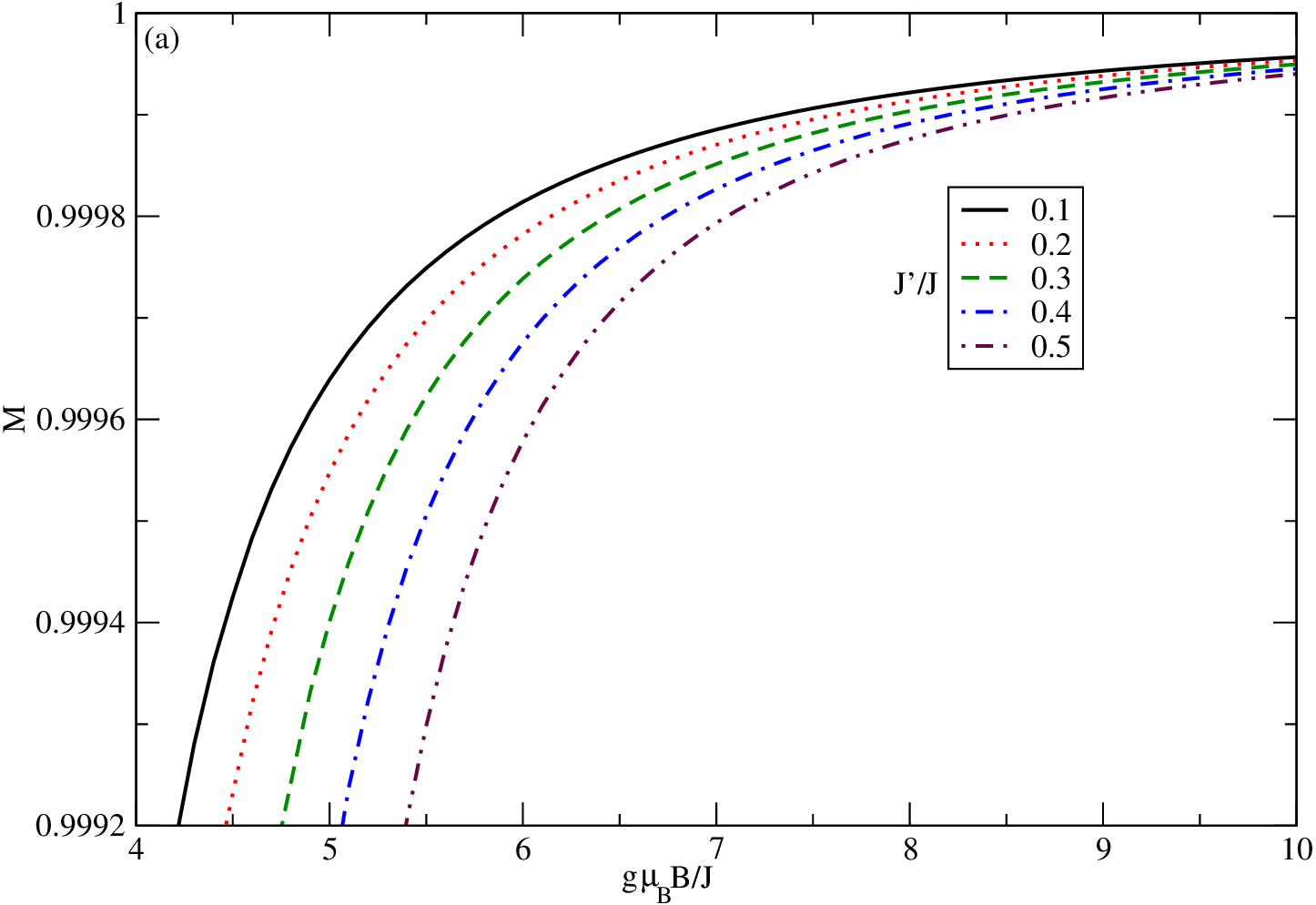}
\includegraphics[width=8cm]{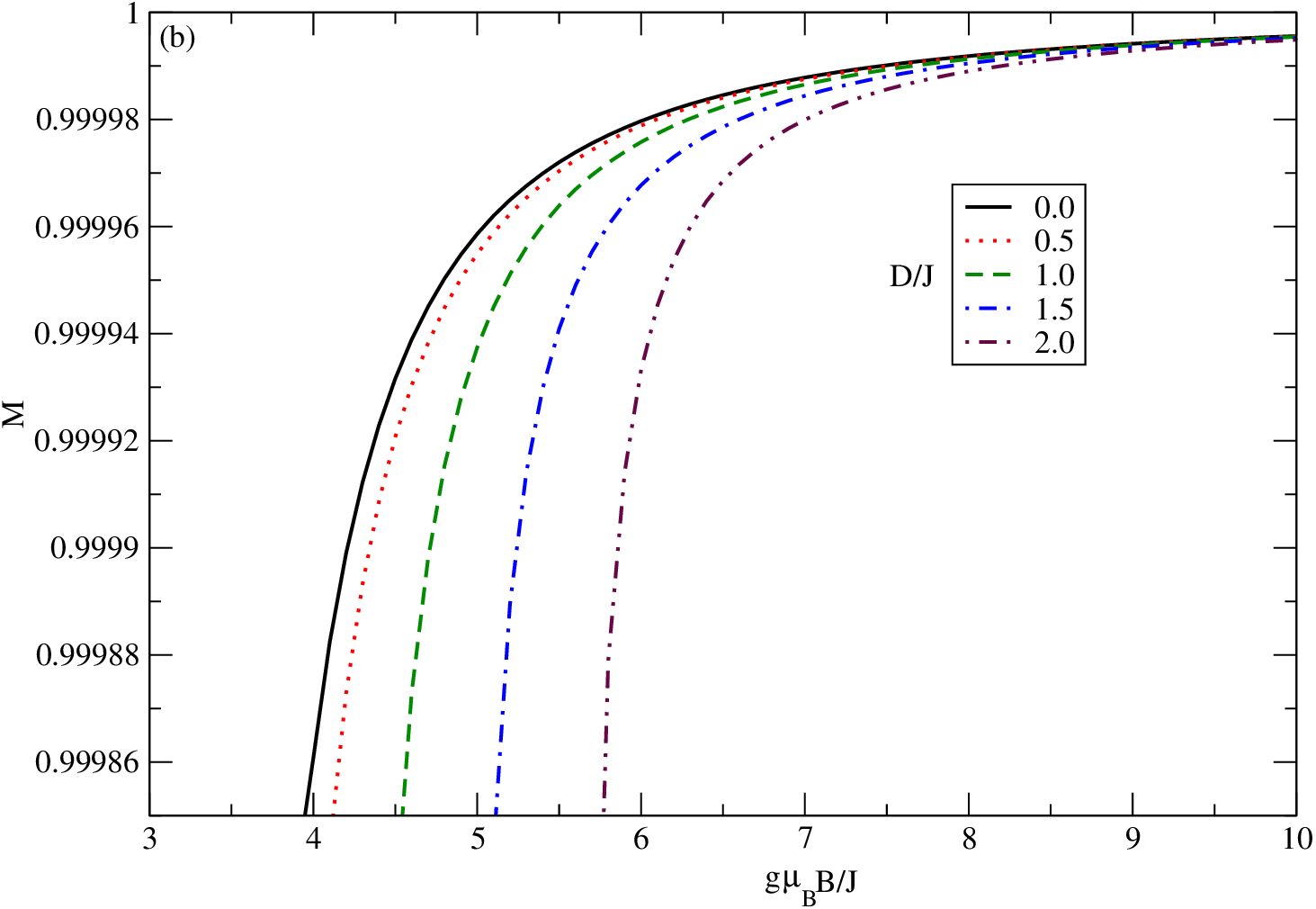}
\caption{(a) Longitudinal magnetization as a function of normalized magnetic field $g\mu_{B}B/J$
for different values of
 next nearest neighbor coupling exchange constant $J'/J$ and
$D/J=0.2$ at fixed temperature and $k_{B}T/J=0.1$.
(b) Longitudinal magnetization as a function of normalized magnetic field $g\mu_{B}B/J$
for different values of
  $D/J$ and
$J'/J=0.2$ at fixed temperature and $k_{B}T/J=0.1$.}
\end{figure}
\begin{figure}
\includegraphics[width=8cm]{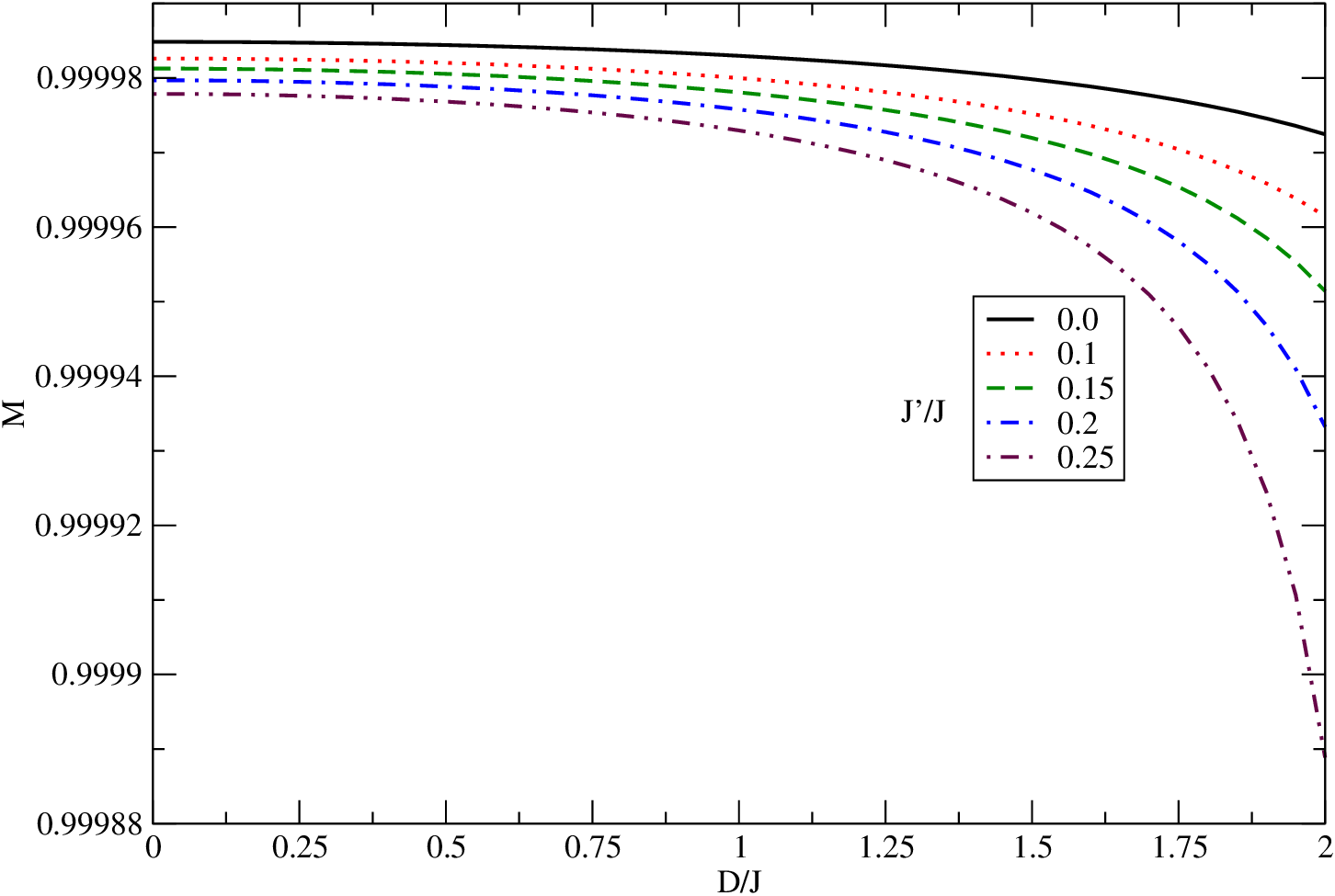}
\caption{Longitudinal magnetization as a function of Dzyaloshinskii-Moriya interaction strength
for fixed normalized magnetic field $g\mu_{B}B/J=6.0$
by setting $k_{B}T/J=0.1$.}
\end{figure}

\section{Appendix: Energy Spectrum and Bogoliuobov coefficients}\setcounter{section}{2}
In this appendix we discuss the details of calculations of excitation energies and Bogoliubov coefficients $u,v$ presented
in Eqs.(\ref{e88888},\ref{e99999}).
In the following,
 we will demonstrate that energy spectrum of non interacting part of model Hamiltonian in Eq.(\ref{e5}) are real
 in spite of $u_{\bf k}$ is a complex variable.
The complex conjugation of unitary transformation in Eq.(\ref{e77777}) is given by
\begin{eqnarray}
 \alpha^{\dag}_{{\bf k}}=v_{{\bf k}}a^{\dag}_{{\bf k}}+u^{*}_{{\bf k}}b^{\dag}_{{\bf k}}\;\;,\;\;
\beta^{\dag}_{{\bf k}}=-u_{{\bf k}}a^{\dag}_{{\bf k}}+v_{{\bf k}}b^{\dag}_{{\bf k}},
\end{eqnarray}
we consider $v_{\bf k}$ is real variable. Using the property of unitary transformation we can write the operators $(a_{{\bf k}},b_{{\bf k}})$
in terms of ($\alpha_{{\bf k}}$,$\beta_{{\bf k}}$) as follows
\begin{eqnarray}
 a_{{\bf k}}=v_{{\bf k}}\alpha_{{\bf k}}-u_{{\bf k}}\beta_{{\bf k}}\;\;,\;\;
b_{{\bf k}}=u^{*}_{{\bf k}}\alpha_{{\bf k}}+v_{{\bf k}}\beta_{{\bf k}},
\end{eqnarray}
By substitution of above transformation into model Hamiltonian
(Eq.(5)) we can rewrite the model Hamiltonian in Eq.(\ref{e5}) of
the manuscript in terms of operators $\alpha,\beta$ as
\begin{eqnarray}
 \mathcal{H}_{bil}&=&\sum_{\bf k}\Big(\mathcal{A}_{{\bf k}}(v^{2}_{\bf k}+|u_{\bf k}|^{2})+\mathcal{B}_{{\bf k}}v_{\bf k}u^{*}_{\bf k}
+\mathcal{B}^{*}_{{\bf k}}v_{\bf k}u_{\bf k}\Big)\alpha^{\dag}_{\bf k}\alpha_{\bf k}\nonumber\\&+&
\sum_{\bf k}\Big(\mathcal{A}_{{\bf k}}(v^{2}_{\bf k}+|u_{\bf k}|^{2})-\mathcal{B}_{{\bf k}}v_{\bf k}u^{*}_{\bf k}
-\mathcal{B}^{*}_{{\bf k}}v_{\bf k}u_{\bf k}\Big)\beta^{\dag}_{\bf k}\beta_{\bf k}\nonumber\\
&+&\sum_{\bf k}\Big((\mathcal{B}_{{\bf k}}v^{2}_{\bf k}-u^{2}_{\bf k}\mathcal{B}^{*}_{{\bf k}})\alpha^{\dag}_{\bf k}\beta_{\bf k}+h.c.\Big).
\label{e555555}
\end{eqnarray}
In order to diagonalize bilinear part of model Hamiltonian in terms
of new operators $\alpha_{\bf k}$ and $\beta_{\bf k}$, we should
apply the following relation
\begin{eqnarray}
 \mathcal{B}_{{\bf k}}v^{2}_{\bf k}-u^{2}_{\bf k}\mathcal{B}^{*}_{{\bf k}}=0.
\label{e11111}
\end{eqnarray}
In other hand the unitary transformation of Eq.(\ref{e77777}) in the manuscript implies
\begin{eqnarray}
 v^{2}_{\bf k}+|u_{\bf k}|^{2}=1.
\label{e22222}
\end{eqnarray}
Using Eqs.(\ref{e11111},\ref{e22222}), we can obtain $u_{\bf k}$
and $v_{\bf k}$ as
\begin{eqnarray}
 u_{\bf k}=\sqrt{\frac{\mathcal{B}_{{\bf k}}}{2\mathcal{B}^{*}_{{\bf k}}}}\;\;,\;\;v_{\bf k}=\sqrt{\frac{1}{2}}.
\label{e3333}
\end{eqnarray}
According to Eq.(\ref{e555555}), two branches of energy spectrum, i.e. $\omega_{\alpha}({\bf k})$ and $\omega_{\alpha}({\bf k})$,
of bilinear part of model Hamiltonian are given by
\begin{eqnarray}
\omega_{\alpha}({\bf k})&=&\mathcal{A}_{{\bf k}}(v^{2}_{\bf k}+|u_{\bf k}|^{2})+\mathcal{B}_{{\bf k}}v_{\bf k}u^{*}_{\bf k}
+\mathcal{B}^{*}_{{\bf k}}v_{\bf k}u_{\bf k},\nonumber\\
\omega_{\beta}({\bf k})&=&\mathcal{A}_{{\bf k}}(v^{2}_{\bf k}+|u_{\bf k}|^{2})-\mathcal{B}_{{\bf k}}v_{\bf k}u^{*}_{\bf k}
-\mathcal{B}^{*}_{{\bf k}}v_{\bf k}u_{\bf k}.
 \label{e444444}
\end{eqnarray}
However $u_{\bf k}$ is a complex variable, Eq.(\ref{e444444})
implies both $\omega_{\alpha}({\bf k})$ and $\omega_{\beta}({\bf
k})$ are real functions and thus Hamiltonian remains as an hermitian
operator. By substitution of Eqs.({\ref{e22222}},\ref{e3333}) into
Eq.(\ref{e444444}), the energy spectrum of non interacting bosons
takes the following relations
\begin{eqnarray}
\omega_{\alpha}({\bf k})&=&\mathcal{A}_{{\bf k}}+|\mathcal{B}_{{\bf k}}|\;\;,\;\;
\omega_{\beta}({\bf k})=\mathcal{A}_{{\bf k}}-|\mathcal{B}_{{\bf k}}|.
 \label{e66666}
\end{eqnarray}

\end{document}